\documentstyle[aps,twocolumn,epsfig]{revtex}

\input epsf.tex

\newcommand{\beq}{\begin{equation}}
\newcommand{\eeq}{\end{equation}}
\newcommand{\beqa}{\begin{eqnarray}}
\newcommand{\eeqa}{\end{eqnarray}}
\def\opone{\leavevmode\hbox{\small1\kern-3.8pt\normalsize1}}

\def\<{\langle}
\def\>{\rangle}

\newcommand{\braket}[2]{\mbox{$ \langle #1 | #2 \rangle $}}
\newcommand{\ket}[1]{\mbox{$ | #1 \rangle $}}

\title{Quantum cryptography with and without entanglement}
\author
{N. Gisin and N. Brunner\\
{\protect\small\em Group of Applied Physics, University of Geneva, 1211 Geneva 4, Switzerland}\\
}
\date{\today}

\begin{document}

\maketitle

\begin{abstract}
Quantum cryptography is reviewed, first using entanglement both
for the intuition and for the experimental realizations. Next, the
implementation is simplified in several steps until it becomes
practical. At this point entanglement has disappeared. This method
can be seen as a lesson of Applied Physics. Finally, security
issues, e.g. photon number splitting attacks, and counter-measures
are discussed.
\end{abstract}


\section{Introduction}
Quantum cryptography is a beautiful idea! It covers aspects from
fundamental quantum physics to Applied Physics via classical and
quantum information theories \cite{RMP}. During the last ten
years, quantum cryptography progressed tremendously, in all
directions: from mathematical security proofs of idealized
scenarii to commercial prototypes. In these proceedings we review
the intuition, the experimental progress in optical fibers
implementations and some security aspects, each viewed first with
entanglement, and then without. Undoubtedly, quantum cryptography
is intellectually more fascinating and conceptually easier with
entanglement, but much more practical without it. Hence both
aspects, with and without entanglement, are equally beautiful!

The next section presents the intuition behind quantum
cryptography. Section \ref{AppliedPhysics} can be seen as a lesson
in Applied Physics: how to simplify a theorist's implementation of
a nice idea until it is practical, while keeping the essential.
This shows that Applied Physics requires a lot of imagination and
a deep understanding of the essential physical ingredients.
Finally, section \ref{security} reviews some security issues:
coherent and individual eavesdropping, Trojan horse attacks,
photon number splitting attacks and means to limit their
efficiency.

\section{Intuitions}\label{intuition}
\subsection{Key distribution}
The general scenario for key distribution, whether classical or
quantum, goes as follows. Alice and Bob, the honest parties, hold
many realizations of random variables X and Y respectively. The
adversary, Eve, holds realizations of a third random variable Z.
Hence the scenario is described by a joint probability
distribution $P(X,Y,Z)$ \cite{Maurer93}. Intuitively it is clear
that if $X$ and $Y$ are strongly correlated (e.g. almost
identical) and  furthermore, if $Z$ is essentially uncorrelated,
then Alice and Bob can use a public communication channel to
distil secret bits. This intuition is made precise in the
following theorem. The useful measure of correlation here is the
mutual Shannon information.

{\em Theorem}\cite{CsizsarKorner78}
For a given $P(X,Y,Z)$, Alice and Bob can establish a secret
key (using only error correction and classical privacy
amplification) if and only if $I(X,Y)\geq \min \{I(X,Z),I(Y,Z)\}$,
where $I(X,Y)=H(X)-H(X|Y)$ denotes the mutual information and $H$
is the Shannon entropy.

Note that by definition {\it privacy amplification} uses only
1-way communication. If Alice and Bob use 2-way communication, the
situation is more complex
\cite{MaurerWolf99,GisinWolf99,AcinGisinMassanes03}. But these
2-way protocols are so inefficient that in practice they are
always ignored.

\subsection{Quantum key distribution with entanglement}
Let us assume that the random variables $X,Y$ and $Z$ introduced
above result from quantum measurements that Alice, Bob and Eve
perform on a quantum state $\psi_{ABE}$. It is clear for the
quantum physicists, that if the partial state $\rho_{AB}$ shared
by Alice and Bob is close to maximally entangled, then Eve is
"factorized out", i.e. is uncorrelated. This is because a
maximally entangled state is a pure state $\rho_{AB}\approx
|\psi_{AB}\rangle\langle\psi_{AB}|$, hence the global state has to
be close to a product state:
$\psi_{ABE}\approx\psi_{AB}\otimes\psi_E$. If one understands
entanglement, more precisely, if one is familiar with the algebra
of tensor products, then the reason why quantum key distribution
with entanglement is secure becomes very intuitive!

\subsection{Quantum key distribution without entanglement}
Assume now that Alice and Bob do not share an entangled state, but
- following the original idea \cite{BB84} - that Alice sends
individual quanta to Bob (when the quanta are described by a
2-dimensional Hilbert space, one speaks of qubits). Alice and Bob
use two (or more) incompatible bases to prepare and measure each
quanta. Because of the use of incompatible bases, there is no way
for Eve to make copies of the flying quanta. Indeed, the
no-cloning theorem guarantees that there is no way to copy an
unknown quanta without perturbing its state \cite{nocloning}. Thus
Alice and Bob can check for the presence of an adversary, Eve, by
comparing a sample of their data: if the data is perfectly
correlated, then Eve did not try to copy it and the remaining data
is safe. Each time Alice and Bob happen to have used the same
basis, their data provides them with a secret bit.

This view of QKD without entanglement can be based on different
aspects of quantum physics, like Heisenberg's uncertainty relation
or that quantum measurements perturb the system. But in the end
all these are based on the linearities of quantum kinematics (the
Hilbert space) and dynamics (Schr\"{o}dinger's equation). And this
linearity is also the basis for entanglement, which appears when
one introduces linear combinations of product states. Hence,
intuitively one feels that both QKD schemes are closely related.

\section{Experiments: a lesson in Applied Physics}\label{AppliedPhysics}
\begin{center}
\begin{figure}
\epsfxsize=8cm \epsfbox{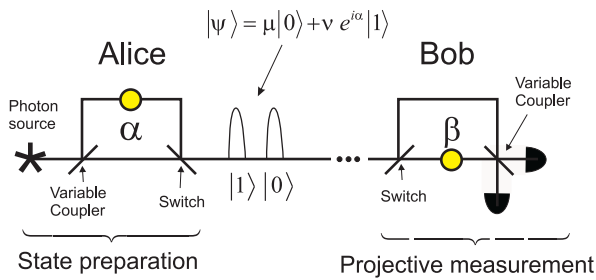} \caption{Time-bin qubits.
After Alice's interferometer the photon is brought into a
superposition of the two time-bins (early and late) corresponding
to the two arms of the interferometer (short and long). The
logical values 0 (1) is attributed to early (late). Note that by
tuning their respective phases and coupling ratios, Alice can
prepare any qubit state and Bob can perform a measurement in any
qubit basis. The switch allows in principle the state preparation
and the measurement without losses. In practice however one often
replaces the switch by a 50-50 coupler and uses postselection. }
\label{timebin}
\end{figure}
\end{center}

The first choice when thinking about an experimental realization
of QKD concerns the degree of freedom used for encoding the qubit.
Indeed, if one goes for optical fibers, then the system is
imposed: telecom photons. A first possible choice would be
polarization. Unfortunately this is a quite unstable degree of
freedom: actual fibers have some birefringence (different
polarization modes travel at different speeds), moreover the
polarization modes suffer from random polarization mode coupling
\cite{PMD}. And if the fiber is hanging between posts, the
situation is even worse: Berry phase would be random, leading to
fast (ms) random polarization fluctuations \cite{BerryPhase}.
Hence, better choices should be envisaged. In Geneva, we chose
time-bin qubits \cite{TimeBinQubit}. The idea is depicted in Fig.
\ref{timebin}. Each photon is brought into a superposition of two
time-bins, an early and a delayed one. The probability amplitudes
of each time-bin and their relative phase allow one to prepare any
possible qubit state. Also any possible projective measurement can
be realized using a similar interferometer shown on the right hand
side of Fig \ref{timebin}.

In the following sub-sections we review step by step
simplifications of the theorist's implementation of QKD.

\subsection{Basic experiment with entangled time-bin qubits}

\begin{center}
\begin{figure}
\epsfxsize=7cm \epsfbox{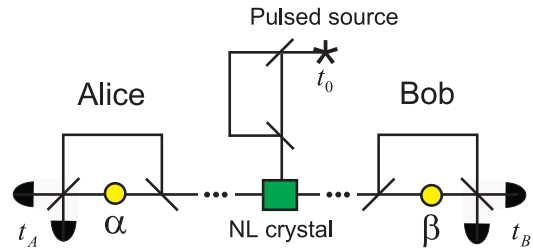} \caption{Quantum cryptography
with entangled time-bin qubits. The source sends a pulse at time
$t_{0}$. The detection on Alice's (Bob's) side occurs at time
$t_{A}$ ($t_{B}$).} \label{tb0}
\end{figure}
\end{center}

\begin{center}
\begin{figure}
\epsfxsize=7cm \epsfbox{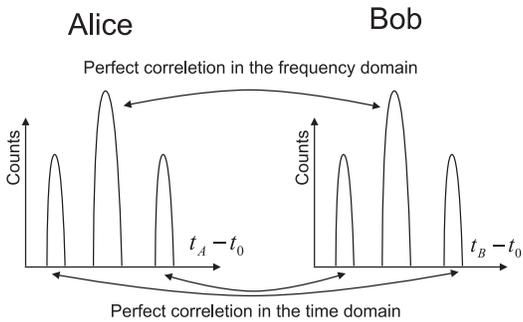} \caption{Time and frequency
correletions} \label{basis}
\end{figure}
\end{center}

The configuration presented in Fig. \ref{tb0} is close to Ekert's
original proposal \cite{Ekert91}, but uses time-bin qubits instead
of polarization. The source at the center contains a non-linear
crystal in which a pump photon spontaneously splits into two twin
photons. Energy conservation guarantees that the twins' energies
(i.e the optical frequencies) add up to the well defined energy of
the pump photon, although each of the twin photon has itself an
uncertain energy, uncertain in the usual quantum mechanical sense.
The pump photon is part of a large classical pulse, about 500 ps
long. Since the probability of "splitting", i.e. of spontaneous
parametric downconversion, is low (typically $10^{-10}$, up to
$10^-6$ in PPLN waveguides \cite{Tanzilli01}), the pulse energy
can be adjusted such that the probability that a pair of twin
photons is generated is around 10\%. In order to produce entangled
time-bin qubits, the pump pulse passes through an unbalanced
interferometer, where the imbalance is much longer than the pulse
duration. Alice and Bob both use the standard time-bin qubit
analyzer presented in Fig. \ref{timebin}. They fix the phases
(relative to the pump interferometer) of their interferometers
such that the two twin photons  always emerge at the same output
port, hence their detectors clicks are perfectly correlated. For a
second, incompatible, basis Alice and Bob could use different
phase settings. But a first simplification can immediately be
implemented. They replace the switch of their measuring
interferometer by a much simpler and less lossy fiber optical
coupler. Hence, Alice and Bob can also detect photons at an
earlier or at a later time: earlier if the pump and their twin
photons passed through the short arms of the interferometers,
later if they both travelled the long way. Consequently, whenever
Alice and Bob both detect their twin photon in the lateral peaks
(i.e. early or late), then they both definitely have the same
detection time: either both early or both late. And if Alice and
Bob both detect their twin photon in the central time-bin, then
they definitely have a click in the same detector (assuming their
phases are fixed at $\alpha=\beta=0$). The first case uses the
time-basis, the second the frequency-basis (see Fig. \ref{basis}).

Conceptually this is quite elegant. It has even been realized in
our lab \cite{Tittel00}, and recent results show that it is
feasible over a significant distance. But this configuration is
not very practical: there are three interferometers to align and
stabilize, and the polarization of the three photons has to be
kept under control. Hence, let's simplify this!

\subsection{First simplification: energy-time entanglement}

\begin{center}
\begin{figure}
\epsfxsize=7.5cm \epsfbox{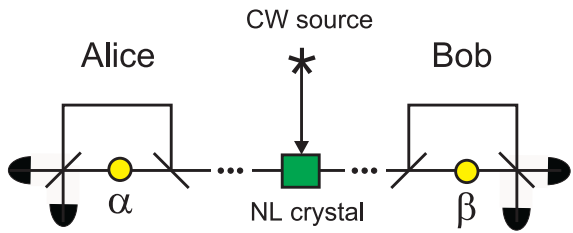} \caption{First simplification.
A continuous wave (cw) source can replace the pump
interferometer.} \label{tb2}
\end{figure}
\end{center}

A first simplification of the previous scheme consists in
suppressing the pump interferometer and replacing the pulsed laser
by a continuous pump laser, see Fig. \ref{tb2}. If the coherence
length of this cw pump laser is larger than the imbalance of Alice
and Bob's interferometers, then 2-photon interference can still be
observed. Indeed, when Alice and Bob post-select coincidence
detections, then there are two possibilities: either both photons
passed through the short arm of both interferometers, or both
passed through the long arm. Since the pump laser's coherence is
large, these two possibilities are indistinguishable. Hence,
according to quantum mechanics, one should add the probability
amplitudes and observe interference. In this configuration Alice
and Bob need to randomly choose the settings of their phase
modulators: 0, 90, 180 and 270 degrees, let's say. Whenever they
happen to use settings corresponding to a phase difference
multiple of $180^o$, then their detectors always fire together.

This is a nice configuration, but admittedly more suited for tests
of Bell inequality (i.e. of quantum non-locality) than for a
practical quantum cryptography setup. Actually, this configuration
has been proposed in 1989 by J. Franson on the context of quantum
nonlocality and is called a Franson interferometer
\cite{Franson89}. This is the configuration we used in 1997 for
our long distance Bell test over 18km in optical fibers (10km in
straight line) \cite{Tittel97}.

\subsection{Somewhat simpler}
The next step notices that there is no need to put the source at
the middle, half-way between Alice and Bob. The middle position is
merely elegant. But it is more practical to put the source on one
side, let's say Alice's side. Notice that Alice doesn't become the
sender of the quantum key: the key results eventually from
independent random choices made by both partners and by Nature,
there is nothing like a quantum key sender. But now, only one
photon must travel a long distance. Hence, the photon that stays
on Alice's side can be chosen at a more convenient wavelength for
efficient detection, that is at a wavelength where silicium APDs
are available, i.e. below 1 $\mu$, around 800 nm. This
configuration, with some additional nice tricks, was demonstrated
in 2001 by G. Ribordy \cite{Ribordy01}, who founded id Quantique a
few years later, the first company to propose a quantum
cryptography setup \cite{idQ}. His experiment was the first one
targeting primarily quantum cryptography with entangled photons -
all other experiments, including ours, where tailored for Bell
tests and merely adapted to fashion. Ribordy's experiment still
holds the distance record of QKD using entangled photons. But
admittedly, is not yet that practical since two photons must be
detected. Hence, let's make it simpler!

\subsection{The first main step towards a practical system}

\begin{center}
\begin{figure}
\epsfxsize=7cm \epsfbox{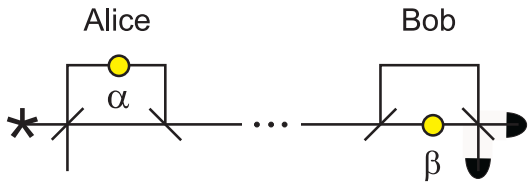} \caption{The source can be moved
on the other side of Alice's interferometer.} \label{tb4}
\end{figure}
\end{center}

The first step towards a really practical system consists in
moving the photon source to the other side of Alice's
interferometer. At first this may look like a complete change, but
it really isn't! Let's first use formulas. Whenever a unitary
operator $U$ acts on one subsystem of a maximally entangled pair
state $\Phi^{(+)}$, then the same effect can be obtained by acting
with a related unitary operator on the other subsystems: \beq
U\otimes \opone \, \Phi^{(+)}=\opone\otimes U^t \, \Phi^{(+)} \eeq
where $U^t$ denotes the transpose.

This formula applied to our case simply tells us that for Alice's
interferometer, the long arm with a central source is equivalent
to the short arm with a source moved to the left of the
interferometer, as shown in Fig. \ref{tb4}. Now the interference
results from the indistinguishability of the following two paths:
short-long and long-short, where the first term applies to the
path in Alice's interferometer and the second to the path in Bob's
interferometer. The significant simplification follows quite
naturally. Since the photon travelling to the left on Fig.
\ref{tb4} is actually not used, or only as a trigger, one may as
well use a single photon source. Well, that is even less
practical, at least as long as single photon sources at telecom
wavelength do not exists. But now one can also use the much more
practical pseudo-single photon sources. These sources are simply
very attenuated telecom laser pulses, such that the mean photon
number per pulse is only of the order of 0.1. Hence the
probability that a pulse contains two photons is almost negligible
(in section III we come back to the issue of multi-photon pulses).
Attenuating a laser pulse that low is not trivial, but still much
simpler and much more stable, which is very important, than
twin-photon sources.

This configuration presented in Fig. \ref{tb4} was first used by
Paul Townsend, then at BT, and John Rarity, then at DERA
\cite{Townsend93}, and is still developed at Los Alamos National
Laboratories, USA, in the group of Richard Hughes \cite{Hugues00}.
But looking at Fig. \ref{tb4} one still sees two interferometers
that need to be stabilized: the difference long-short has to be
the same for both interferometers. And since this scheme relies on
interferences, the polarization of the pseudo-single photons must
be controlled. All this requires active feedback, which is not
impossible to achieve, but not yet entirely practical. So let's
simplify it further!

\subsection{A practical setup: the Plug \& Play configuration}

\begin{center}
\begin{figure}
\epsfxsize=7.5cm \epsfbox{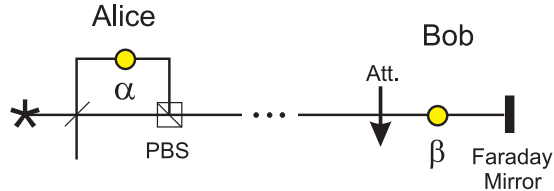} \caption{Plug \& Play setup
[23,24].} \label{tb5}
\end{figure}
\end{center}

The next step realizes that there is no need for two
interferometers, one is enough (see Fig. \ref{tb5}). But then the
pulse must travel go-\&-return, using a mirror as indicated on the
figure. The indistinguishable paths are still short-long and
long-short, but now referring to the paths during the go and the
return propagations. Notice that in this scheme the role of Alice
and Bob are inverted: Bob chooses one among four phase settings
and Alice chooses a measurement basis. A serious drawback is that
the photons must travel twice the distance, hence suffer from
twice the loss. But this can be circumvented. Actually it is only
on the return flight that the pulse has to be attenuated down to
the pseudo-single photon level. Consequently, a bright pulse is
sent out, attenuated by Bob and reflected to Alice. Notice that
since there is now only a single interferometer, there is no
longer any need to align it! All that is needed is that it remains
stable during the time of a go-\&-return, i.e. a few
micro-seconds. But there remains the polarization. Here again
there is an elegant solution, first suggested in a different
context by Martinelli \cite{Martinelli89}. It consists of using a
Faraday mirror. The details can be found in
\cite{Martinelli89,RMP}. Essentially such mirrors act on
polarization like a phase conjugating mirror acts on phase. The
net result is that when a light pulse arrives back on Alice's
side, it is in a fixed polarization state, independent of all the
polarization fluctuation light underwent during propagation: all
the fluctuations where undone during the return journey. Faraday
mirrors use the non-reciprocal Faraday effect, the same effect
used in isolators and in circulators. Hence the telecom industry
has developed this technology to a remarkable point and Faraday
mirrors can readily be bought \cite{JDS}.

A further simplification comes from the fact that a Faraday mirror
exchanges vertical and horizontal polarization. Hence, replacing
the output coupler of Alice's interferometer by a polarization
beam splitter guarantees that a photon that passed through the
long arm when emitted, will return via the short arm, and
vice-versa. Consequently, Alice doesn't need to post-select the
cases where the photon arrives in the correct time-bin since all
detected photons arrive at the correct time.

When this setup was first tested using classical light (i.e.
without the attenuator), experimentalists in Geneva were very
pleased to measure visibilities up to $V=99.8$\%, without much effort,
even over tens of km! Accordingly this configuration was named
Plug-\&-Play \cite{Muller97}. Using this setup for QKD, the noise (QBER) is
largely dominated by the detector noise: $QBER_{optical}=\frac{1+V}{2}<<1\%$.

The Plug \& Play configuration has been demonstrated in a QKD
experiment between Geneva and Lausanne over a distance of 67km
using the swiss telecom network, with terrestrial and with a cable
under lake Geneva \cite{Stucki02} (see also \cite{BethuneRisk00}).
This experiment received quite a lot of attention. But actually,
another experiment presented in the same paper deserves probably
more attention. It used aerial cables and was done in mountains
near Geneva. This clearly demonstrated the very high stability of
the Plug \& Play configuration. Indeed, it would be almost
impossible to demonstrate QKD with any of the previously discussed
configurations using aerial cables!

\section{Security}\label{security}

\subsection{Security proofs based on entanglement}
The most general proofs of security, often termed \`{a} la
Shor-Preskill, are quite surprising \cite{ShorPreskill00}.
Following ideas by Mayer \cite{Mayers98}, Lo and Chau
\cite{LoChau99} and the development of quantum error codes, these
proofs essentially show that from Alice and Bob's points of view
everything is as if they had used close to maximally entangled
states, although they actually did use a scheme without
entanglement. More details can be found in I. Chuang's
contribution to these proceedings. Let us simply emphasize that it
is still not known whether Eve can in principle reach these
bounds, or whether these bounds are sub-optimal. From a practical
point of view this is a pity, since we do not know whether we do
really need to sacrifice qubits to these bounds or could use the
more optimistic bound summarized in the next sub-section.

\subsection{Security proofs without entanglement}
The proofs in this subsection do not consider the most general
attack, but only what is called the individual, or incoherent
attack. Actually, these proofs also treat the case of
finite-coherent attacks, hence let us concentrate on the later.
All the security proofs are valid only in the limit of arbitrarily
long keys. If not, the statistical arguments wouldn't apply. Now,
let's assume that Eve can attack several qubits in a coherent way,
i.e. she can coherently let auxiliary systems under her control
interact (unitarily of course) with the flying qubits. Assume that
Eve can do this up to a maximum number of N qubits. We call this
finite-coherent attacks. If Alice and Bob use key lengths much
longer than N, then Eve is in the same situation as if she would
be limited to individual attacks (one auxiliary system per qubit).
Hence, the proofs "without entanglement" are valid for all senarii
except if Eve can attack coherently an unlimited number of qubits
- a conceptually interesting scenario, but hard to take seriously
for the practical physicist.

\begin{center}
\begin{figure}
\epsfxsize=7.5cm \epsfbox{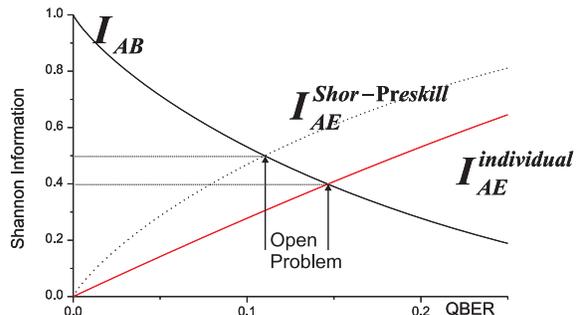} \caption{The Shannon
mutual information as a function of the QBER for the BB84
protocol. It is still not known if the Shor-Preskill bound
(QBER$\approx$11 \%) can be saturated or not. The bound for
invidual attacks (QBER$\approx$15\%) is known to be optimal.}
\label{information}
\end{figure}
\end{center}

In 1997 Fuchs et al. \cite{Fuchsetal97} presented the optimal
individual attack, see also \cite{Lutkenhaus00}. Since then it has
been generalized to more than two bases \cite{BechmannGisin99} and
to higher dimensions \cite{HighDim}. By now, the BB84 case is well
known. The main results are summarized in Fig. \ref{information}.

\subsection{Trojan horse attacks and technological loopholes}

As shown in Fig. \ref{tb5} the configuration opens a new possible
attack for Eve, the so-called Trojan horse attack. Eve could send
into Bob's apparatus a bright laser pulse to sense the phase
modulator's setting. This illustrates that for every
simplification step one has to carefully check the security of the
configuration. In the present case there is a simple way to avoid
Trojan horse attacks. Bob adds a coupler taking out a large
fraction (typically 90\%) of the light at his apparatus input.
This coupler can be considered as part of the attenuator shown in
Fig. \ref{tb5}. The extracted light is directed onto a standard
detector that monitors the energy of each incoming pulse.
Additionally this detector is very useful for the synchronization
of the phase modulator. Eve could now use a different wavelength
at which either the coupler or the detector is inefficient. To
avoid this Bob has to use a filter which blocks all unwanted
wavelengths. This discussion could be extended more or less for
ever. Let us emphasize two important points. First, this is not
specific to the Plug-\&-Play configuration, every real optical
component has some imperfection, in particular they do all reflect
some light. Hence Eve could always try to send a sensing pulse and
Alice and Bob should always have warning detectors and protecting
filters. The second point is that this brief discussion
illustrates the limit of mathematical proofs of security. Indeed,
such proofs have either to assume perfect components, or
components with precisely defined defects. In practice a central
issue is how to make sure that an actual prototype satisfies the
assumptions of a mathematical theorem? In this respect, it should
be mentioned that when we made the simplification from a 2-photon
to a 1-photon configurations, we lost the possibility of using the
violation of Bell's inequality as a signature of quantumness (i.e.
if the correlation measured by Alice and Bob violate some Bell
inequality, then they definitely share an entanglement preserving
quantum channel). Using Bell inequalities in this sense is a very
nice idea \cite{MayersYao98}. However the detection efficiency
loophole that affects all optical tests of Bell inequality renders
this kind of control infeasible with near future technology
\cite{GisinGisin99}. Note also that a violation of a Bell
inequality could not detect a Trojan horse type of attack.

\subsection{What is secure?}
Since there is some controversy on this, let us ask "what is
secure in QKD?". It is clear that Eve should not have access to
Alice nor to Bob's electronics. Indeed, there the information is
classical and Eve could merely copy it. On the contrary, the
quantum channel, i.e. the optical fiber, is secure thanks to
quantum physics. But now comes an old question in a new context:
where does the quantum/classical transition happen? As long as the
information is quantum, the no-cloning theorem applies. As soon as
it is classical, security is lost (i.e. must be guaranteed by
other means). Surprisingly to us, many physicists (mainly
theorists) consider the detector on the quantum side. This is of
course a simple way to be on the safe side \cite{L}. But it
implies a very significant waste of qubits. It seems really hard
to imagine Eve modifying Bob's detector's dark count probability
from a distance. And if we give her this capability, why not also
give her the power to change Alice's source from a distance? Let's
say that quantum cryptography offers "only" secure key
distribution over a quantum channel, assuming the hardware on both
sides are secured by classical means.

There remains though an issue. Eve could modify the apparent
detection efficiency of Bob's detector by sending brighter pulses.
This is clearly feasible and Bob thus has to continuously monitor
the coincidence rate between his detectors. If this coincidence
rate exceeds the threshold corresponding to accidentals (due
mainly to dark counts), then he should interrupt the protocol.

\subsection{Multi-photon pulses: problem and solutions}
Another potential security loophole comes from the cases where the
pseudo-single photon source actually produces more than one
photon. These events being rare one may think that they are
negligible. However, if the losses on the quantum channel are
high, e.g. the fiber is long, then the cases where the desired
photon makes it to Bob are also rare. Hence Eve could perform the
following attack \cite{BrassardLutkenhausetal00}. Directly at the
exit of Alice's office, Eve counts the number of photons in each
pulse, without perturbing the degree of freedom used to encode the
qubit, i.e. Eve performs quantum nondemolition measurements on
each pulse (this is total science fiction with today's technology,
but if one assumes that Eve is limited only by the laws of
physics, she could do so). Next, Eve blocks all single-photon
pulses. Whenever a pulse contains 2 or more photons, she keeps one
and sends the others to Bob through a perfect channel, or even
better she teleports them to Bob. If the fraction of pulses Eve
blocks balanced the fraction of pulses that would have got lost in
normal operation, then Bob notices no difference. But now Eve
holds a copy of the qubits. The main point is that she didn't need
to make any copy, Alice unwillingly offered her some.

Of course, once the attack was performed, Eve has to conserve her
photons, waiting for the basis reconciliation, when Alice publicly
announces which basis she used to encode each qubit. Thus Eve
clearly needs a quantum memory, which again is far from today's
technology but could in principle be designed.

A first way around such PNS (Photon Number Splitting) attack
consists of using sources producing sub-poissonian light. Indeed
in such sources, the probability of 2-photon pulses is reduced
compared to a poissonian light source like a laser, for the same
probability of a 1-photon pulse. Such sources are often named
single-photon sources and are an active field of research
\cite{Yamamoto1photon}.

\begin{center}
\begin{figure}
\epsfxsize=5.5cm \epsfbox{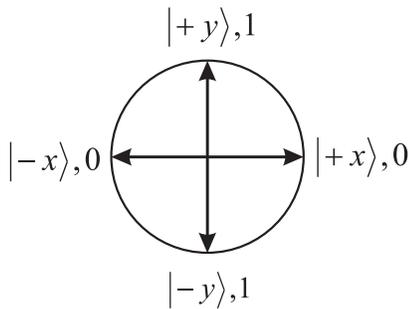} \caption{The SARG protocol.
The hardware is exactly the same as for the well known BB84
protocol. The only difference is in the sifting procedure. To
exchange a secret bit Alice and Bob proceed as follows. Alice
prepares one of the 4 states shown above, say $\ket{+x}$. Then she
announces to Bob a set of two non-orthogonal states containing the
state she actually prepared, for example $\{ \ket{+x},\ket{+y}\}$.
Whenever Bob measures in the $x$ basis (the {\em correct} basis),
he always finds $\ket{+x}$ and cannot conclude anything. But when
Bob measures in the $y$ (the {\em wrong} basis) he finds
$\ket{-y}$ half of the time. In these cases he concludes that
Alice prepared the state $\ket{+x}$. The PNS attack is much less
effective for this protocol than for the BB84, since Eve has to
distinguish between two non orthogonal states. The probability of
success of such a measurement is $p=1-\gamma\approx0.29$, where
$\gamma=\braket{\pm x}{\pm y}=\frac{1}{\sqrt{2}}$ since we used
two maximally conjugated basis. } \label{sarg}
\end{figure}
\end{center}

Another approach realizes that the weakness of the BB84 protocol
against PNS attacks is that whenever Eve holds a copy, she has
full information about the quantum state. But then, why not
replace in the protocol the bases by sets of non-orthogonal states
\cite{SARGPRA}. Remember that unambiguous discrimination of non
orthogonal states is possible, but at the cost of some
inconclusive results \cite{PeresBook}. So even when Eve has a
perfect copy of the state she cannot find out what the bit is with
certainty. A particularly simple example of such new protocols
uses precisely the same states and measurements as in the BB84
protocol, but the sifting procedure differs \cite{SARGPRL}. This
protocol is called SARG and is described in Fig.\ref{sarg}

Though PNS attacks seem completely unrealistic with today's
technology, it is nice to see that new protocols can still be
devised, inspired by practical considerations. In this respect,
see also Ph. Grangier's contribution to these proceedings.

\section{Conclusion}
Quantum cryptography is a beautiful idea! It is also an excellent
teaching tool, encompassing basic quantum physics (no-cloning
theorem, entanglement) and Applied Physics (telecom engineering).
It also involves a significant part of classical and of quantum
information theory.

\small
\section*{Acknowledgements}
Financial support by the Swiss OFES within the European project
RESQ and the NCCR {\it Quantum Photonics} are acknowledged. Many
thanks to Rob Thew for careful reading of the manuscript.

\end{document}